\documentclass[12pt]{article}

\usepackage{graphicx, bm, amsmath}% amssymb, amstext, amsfonts
\usepackage{floatflt}
\begin{document}
%\titlerunning{Title running}
\begin{center}
{\Large\bf \boldmath The Glueball Spectrum from Constituent Models} %<== title (bold face, capitalize)

\vspace*{6mm}
{V. Mathieu$^a$, F. Buisseret$^a$, C. Semay$^a$ and B. Silvestre-Brac$^b$ }\\      %<== authors
{\small \it $^a$ Universit\'e de Mons-Hainaut, Acad\'{e}mie universitaire Wallonie-Bruxelles \\      %<== institutions
            $^b$ LPSC Universit\'{e} Joseph Fourier, Grenoble 1 \\
CNRS/IN2P3, Institut Polytechnique de Grenoble, 53 Avenue des Martyrs, FR-38026
Grenoble-Cedex, France}
\end{center}

\vspace*{6mm}

% abstract
\begin{abstract}
We present a model for odd-$C$ (negative charge parity) glueballs with three constituent
gluons. The model is an extension of a previous study of two-gluon glueballs. We show that,
even if spin-1 gluons seem to reproduce properly the lattice QCD spectrum for $C=+$ states,
the extension for $C=-$ cannot match with the lattice results. Resorting to the helicity
formalism, we show how transverse gluons fit in better agreement the lattice QCD spectrum.
\end{abstract}

\vspace*{6mm}

\section{Constituent models for two-gluon glueballs}
Quantum Chromodynamics (QCD) allows the self-coupling of the gauge bosons, the gluons.
Therefore, states with no valence quarks, the glueballs, are a beautiful consequence and
prediction of QCD. Recently, a comprehensive review was devoted to the glueballs
\cite{Mathieu:2008me}.

Their observation, however, remains difficult. Probably because the lightest glueball, the
scalar $0^{++}$, should mix with mesons~\cite{Klempt}. Some experimental glueball candidates
are currently known, such as the $f_0(980)$, $f_0(1500)$, $f_0(1710)$, \dots but no definitive
conclusions can be drawn concerning the nature of these states.

On the other hand, pure gauge QCD has been investigated by lattice QCD for many years, leading
to a well established glueball spectrum below 4~GeV \cite{lat0,lat1,lat3}. Our aim is to
reproduce this hierarchy with the most simple models with constituent gluons. Since two gluons
can only bind into positive-$C$, we have to consider three-gluon glueballs for the existence
of negative-$C$ states.

In ref.~\cite{brau}, the authors provide a relevant model of two-gluon glueballs. Assuming
Casimir scaling for the string tension of the flux tube, the Hamiltonian, endowed with
one-gluon exchange (OGE) potentials, reads
\begin{equation}\label{Hgg1}
    H_{gg} = 2\sqrt{\bm p^2 + m^2} + \frac{9}{4}\sigma r + V_{oge}(r; \alpha_S,\mu; \bm S, \bm L).
\end{equation}
Although they use a bare mass $m=0$ in the kinetic term, their gluons have longitudinal
components and are spin-1 particles. Therefore, many states are degenerate and the authors
resorted to spin-dependent potentials coming from the OGE to lift these degeneracies. The
corrections are of order $\mu^{-2}$, where $\mu=\left\langle\bm p^2\right\rangle$ is an
effective constituent mass. The parameters were fitted on the low-lying states and the final
spectrum is displayed in Fig.~\ref{fig1}~(left).

All states (squares) fall into lattice error bars. However, we noticed some spurious states
(circles) not found by any lattice study. $J=1$ states are forbidden by Yang's theorem and
should not be present. The appearance of such states is induced by the longitudinal components
of gluons and should disappear when considering transverse gluons.

\section{Odd-$C$ glueballs}
Let us forget about the spurious states for the moment and let us generalize the model for
three-gluon glueballs. We used a generalisation of the flux tube for the confinement. In heavy
baryons, the confinement has a Y-shape, but in our case, we replaced it by a center-of-mass
junction. The Hamiltonian is supplemented by the potential coming from the OGE and reads
\begin{equation}\label{Hggg}
    H_{ggg} = \sum_{i}^3 \sqrt{\bm p_i^2} + \frac{9}{4}f \sigma |\bm r_i- \bm R_{cm}| +
    \sum_{i<j} V_{oge}(r_{ij}; \alpha_S, \mu; \bm L_{ij}, \bm S_{ij}).
\end{equation}
We refer the reader to the ref.~\cite{Mathieu:2008pb} for further details concerning the
Hamiltonian.

\begin{figure}[h]
\centerline{\includegraphics[width=0.5\linewidth]{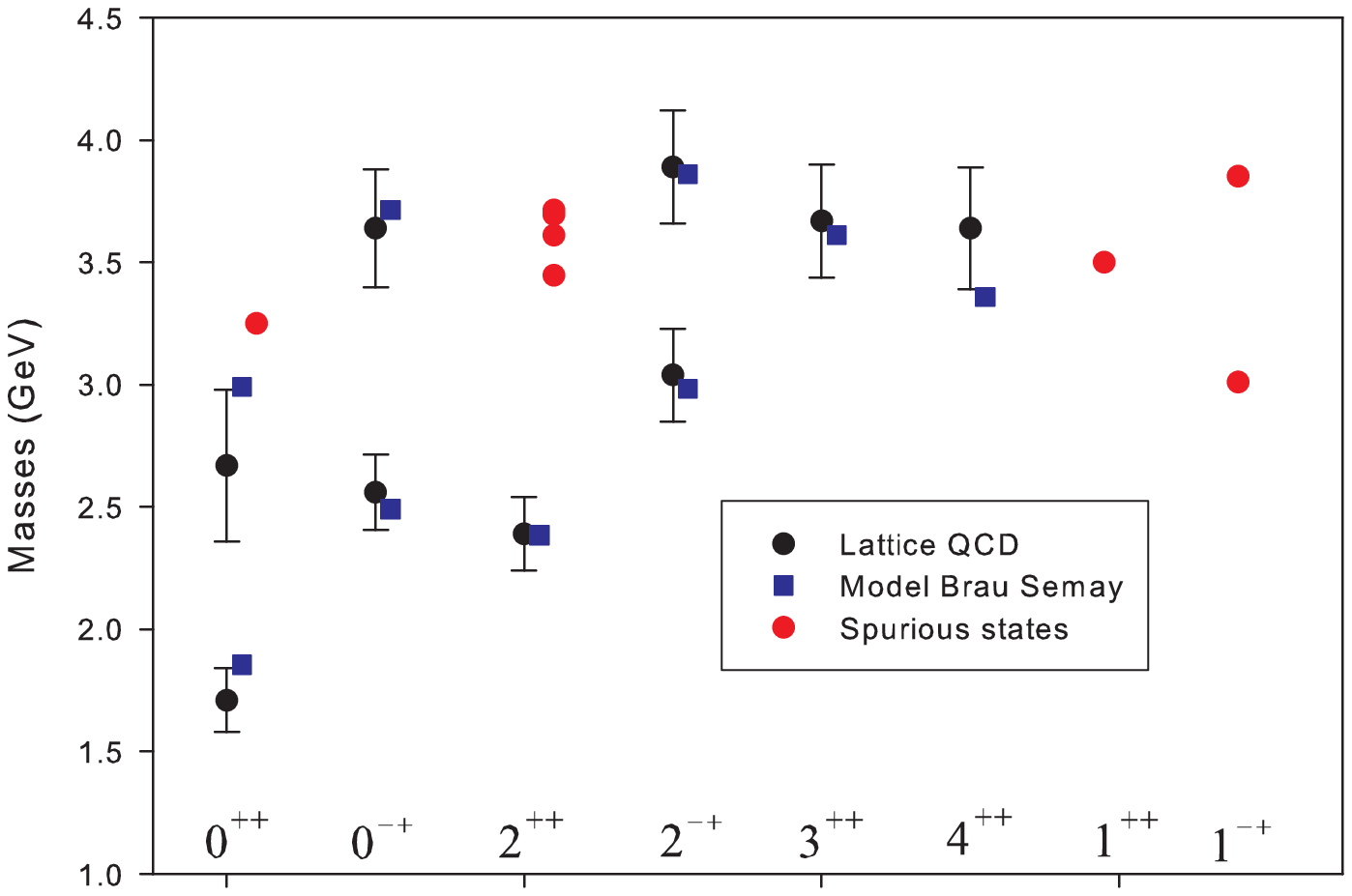}
\includegraphics[width=0.5\linewidth]{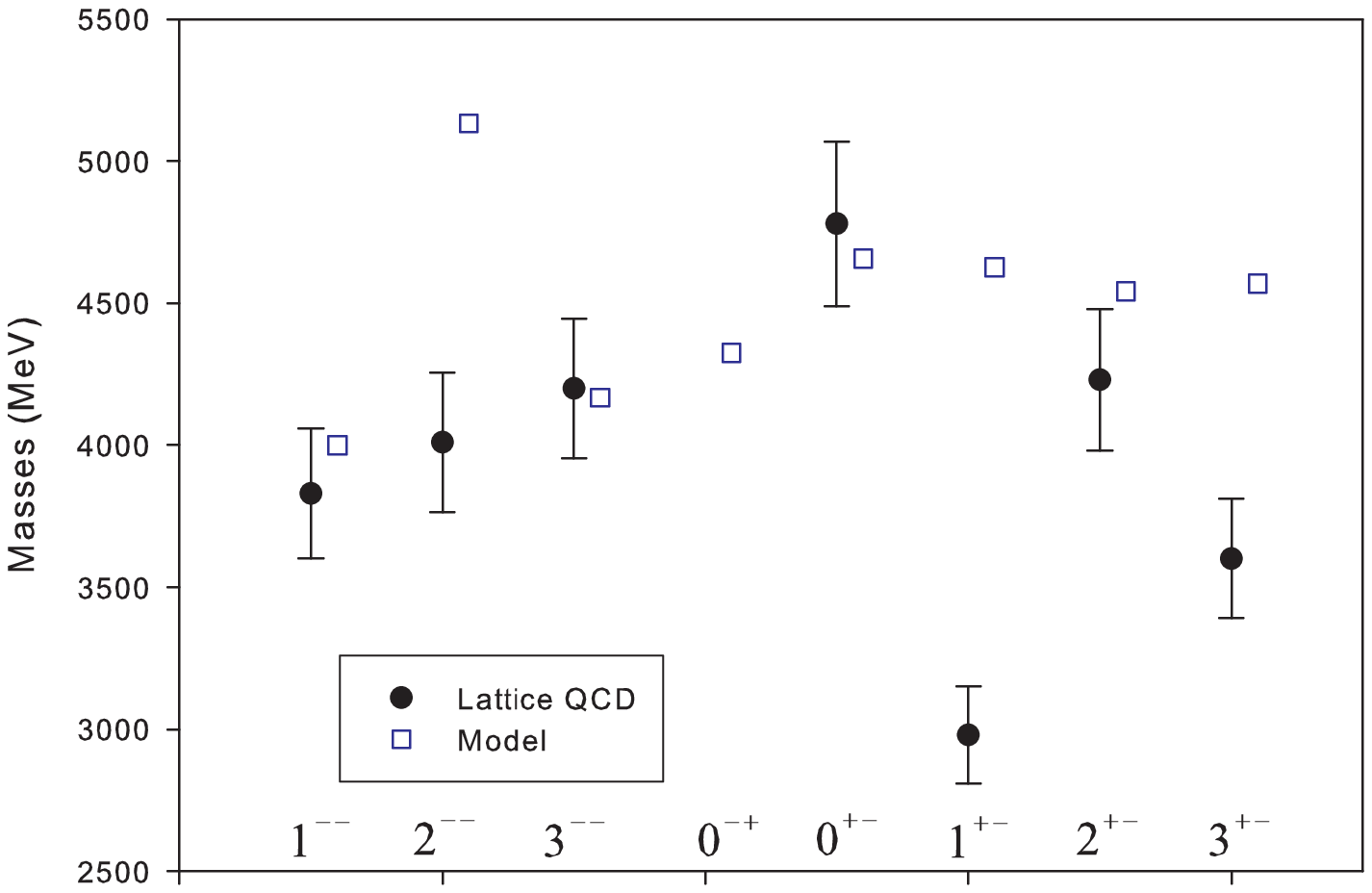}}
\caption{\label{fig1}Left: Spectrum of Hamiltonian~(1) with longitudinal gluons. Right:
Spectrum of Hamiltonian~(2) with longitudinal gluons.}
\end{figure}

We impose the symmetric colour function $d_{abc}A^a_\mu A^b_\nu A^c_\rho$, which ensures a
negative $C$-parity, then the spin symmetry determines the symmetry of the space. Since $\bm
1\otimes\bm1\otimes\bm1=\bm3_s\oplus\bm2_m\oplus\bm1_s\oplus\bm0_a$, $2^{--}$ has a mixed
symmetry and cannot lie in the same mass range as $1^{--}$ and $3^{--}$, as was already
noticed in ref.~\cite{Mathieu:2006bp}. Moreover, a positive parity requires an odd angular
momentum. Then, all $(0,1,2,3)^{+-}$ are degenerate with a large component $L=1$ in the wave
function. But the lattice QCD exhibits a gap around 2~GeV between the highest $0^{+-}$ and the
lowest $1^{+-}$. The spectrum, shown in Fig.~\ref{fig1} (right), is nearly in complete
disagreement with lattice QCD. The symmetry arguments are Hamiltonian-independent and we can
therefore conclude that models with longitudinal gluons are not appropriate to reproduce the
lattice pure gauge spectrum.

\section{Transverse gluons}
\begin{floatingfigure}[r]{7cm}
\centerline{\includegraphics[width=0.45\linewidth]{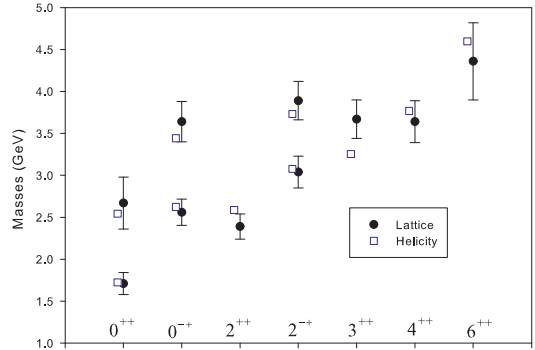}} \caption{Spectrum of
Hamiltonian~(3) with transverse gluons\label{fig2}.}
\end{floatingfigure}

In order to solve the problems encountered (spurious states, hierarchy in the $PC=+-$ sector),
we implemented a formalism developed by Jacob and Wick~\cite{jaco}. This formalism allows us
to handle transverse particles. When applying it to two-gluon glueballs, we remarked that the
Bose symmetry (and the parity) implies selection rules. Three families were
identified~\cite{Mathieu:2008bf}: $(2k)^{++}, (2k+3)^{++},(2k+2)^{-+}$ with $k\in N$.
%(2k)^{++}, && (2k+3)^{++}, && %\\
%(2k+2)^{-+},&& k\in N .%\mathds{N}
%\end{align}
One easily checks that no spurious $J=1$ states appear. Moreover, with this special
construction, all states are now expressed through a given linear combination of spectroscopic
states $|^{2S+1}L_J\rangle$. The degeneracies occurring in ref.~\cite{brau} are naturally
split by the wave function. One does not need to use complicated spin-dependent potentials.

We tested the wave functions with a simple Hamiltonian:
\begin{equation}\label{Hggg}
    H_{gg} = 2 \sqrt{\bm p^2} + \frac{9}{4}\sigma r - 3\frac{\alpha_S}{r}.
\end{equation}
The resulting spectrum, displayed in Fig.~\ref{fig2} is in good agreement with the lattice QCD
data without the inclusion of spin-dependent potentials. But instanton-induced interactions
were needed for $J=0$ states.  In addition, all states are present with no spurious state.

The next step is to implement this formalism for three-gluon glueballs. This work is under
construction. However, we have some indications that the lowest odd-$C$ are spin 1 and
3~\cite{Boulanger:2008aj}. Symmetry arguments are also in favour of a four-gluon
interpretation for $0^{+-}$.

\section*{Acknowledgement}
V. Mathieu thanks the organizers the invitation to the conference and the Dubna Joint
Institute for nuclear research for its hospitality.

\end{document}